\documentstyle[epsf,floats,prd,aps]{revtex}

\begin{document}
\draft

\twocolumn[\hsize\textwidth\columnwidth\hsize\csname @twocolumnfalse\endcsname
\title{Testing a Simplified Version of Einstein's Equations
for Numerical Relativity}

\author{Gregory B.\ Cook}
\address{Center for Radiophysics and Space Research,
		Cornell University, Ithaca, New York\ \ 14853}
\author{Stuart L.\ Shapiro}
\address{Center for Astrophysics and Relativity, 326 Siena Drive,
	Ithaca, NY\ \ 14850}
\author{Saul A.\ Teukolsky\thanks{Departments
	of Physics and Astronomy, Cornell University}}
\address{Center for Radiophysics and Space Research,
		Cornell University, Ithaca, New York\ \ 14853}

\date{\today}

\maketitle

\begin{abstract}
\widetext
Solving dynamical problems in general relativity requires the full
machinery of numerical relativity.  Wilson has proposed a simpler but
approximate scheme for systems near equilibrium, like binary neutron
stars.  We test the scheme on isolated, rapidly rotating, relativistic
stars. Since these objects are in equilibrium, it is crucial that the
approximation work well if we are to believe its predictions for more
complicated systems like binaries. Our results are very encouraging.
\end{abstract}
\vskip2pc]

\narrowtext
\section{Introduction}
\label{sec:intro}
Some of the most interesting unsolved problems in general relativity
require full dynamical solutions of Einstein's equations in three
spatial dimensions. Such solutions have to be found numerically, and
this is only barely becoming technically feasible.  An important set
of problems in this category is the binary coalescence of black holes
and the binary coalescence of neutron stars. Such events are expected
to be a significant source of gravitational waves that will be
detectable by new generations of detectors such as LIGO.

In Newtonian physics, binary stars can orbit in an equilibrium system.
In general relativity, by contrast, a binary system loses energy by
gravitational wave emission. The orbit shrinks, and the two stars
ultimately coalesce. Though this is clearly not an equilibrium
situation, the orbital decay occurs on a much longer timescale than an
orbital period, at least up until the last plunging orbit when the
stars are very close.  Preliminary calculations of binary coalescence
and gravitational collapse suggest that the amount of energy radiated
gravitationally is small. Thus, even when the system becomes highly
dynamical and far from equilibrium, one might expect that it is still
the nonradiative part of the gravitational field that controls the
evolution.

Wilson\cite{Frontiers:Wilson,Texas_90:Wilson} has proposed an
approximation scheme that tracks the evolution of coalescing binary
neutron stars without solving the full dynamical Einstein field
equations. The method may also be applicable to binary black hole
systems\cite{Note1}.  The scheme applies to systems that are either in
or near equilibrium, in which case a reduced set of Einstein's
equation should adequately describe the system.  For example, a binary
system is near equilibrium as long as the emission of gravitational
radiation is small. In strict equilibrium, as in the case of a single
rotating star, there is a coordinate frame in which the first and
second time derivatives of the metric are zero.  In the $3+1$
formalism, this means in particular that the time derivatives of the
3-metric $\gamma_{ij}$ and the extrinsic curvature $K_{ij}$ are
zero. In quasi-equilibrium, the time derivatives are small, and the
metric and extrinsic curvature will not depart significantly from
their initial values. Wilson's approximation consists in setting time
derivatives exactly equal to zero in a selected subset of Einstein's
equations, and ignoring the remaining dynamical equations. This
approximation results in a smaller, more tractable set of field
equations.  Part of the strategy for selecting the subset of
Einstein's equations is to guarantee that $\gamma_{ij}$ and $K_{ij}$
are solutions of the initial-value (or constraint) equations.  Wilson
has proposed evolving the system through a sequence of initial-value
problems by solving the full dynamical equations for the {\it matter}
in the instantaneous background metric, and then updating the metric
quantities at each time step by re-solving the selected subset of
Einstein's equations. We will outline below a simpler method to track
the evolution, which exploits the near equilibrium of the matter as
well.

As compelling as this idea sounds, it is impossible to calibrate the
approximation without comparing it with solutions to the exact
equations.  No such exact solutions exist for realistic, dynamical
3-dimensional cases.  Only recently has it become possible to solve
Einstein's equations numerically for interesting 2-dimensional
problems. In fact, it is only in the last few years that as simple a
problem as the equilibrium structure of a rapidly rotating
relativistic star could be thoroughly investigated.  In this paper, we
use these rotating equilibrium solutions to calibrate Wilson's
approximation scheme. This is the simplest case for which the
approximation scheme is different from the exact equations. Because
the system is a true equilibrium, it is clearly necessary that the
approximation work well in this case.  Only then will we have
confidence that the method is at all useful in more complicated
situations such as binary systems.

\section{Basic equations}
\label{sec:basic_eqns}
A general metric may be written in $3+1$ form as
\begin{equation}
ds^2 = - \alpha^2 dt^2 + \gamma_{ij} (dx^i + \beta^i dt) (dx^j + \beta^j dt).
\end{equation}
The dynamical equation for $\gamma_{ij}$ is
\begin{equation}
\label{eqn:met_evol}
\partial_t \gamma_{ij} = -2\alpha K_{ij} + D_i \beta_j + D_j \beta_i,
\end{equation}
where $D_i$ denotes a covariant derivative with respect to
$\gamma_{ij}$.  The trace of this equation is
\begin{equation}
\partial_t \ln \gamma^{1/2} = - \alpha K + D_i \beta^i,
\end{equation}
where $\gamma=\hbox{det}\gamma_{ij}$ and $K=K^i{}_i$. The trace-free
part of Eq.~(\ref{eqn:met_evol}) is
\begin{eqnarray}
\label{eqn:met_evol_tf}
\gamma^{1/3}\partial_t(\gamma^{-1/3} \gamma_{ij}) &=&
-2\alpha (K_{ij} - {1 \over 3} \gamma_{ij} K) + \nonumber \\
&&\mbox{} + D_i \beta_j + D_j \beta_i -
{2 \over 3} \gamma_{ij} D_k \beta^k.
\end{eqnarray}
We fix the six components of the extrinsic curvature $K_{ij}$ by
demanding that each data slice be a maximal slice and that the
left-hand side of Eq.~(\ref{eqn:met_evol_tf}) be equal to zero. This
gives
\begin{equation}
\label{eqn:max_slice}
K = 0,
\end{equation}
and
\begin{equation}
\label{eqn:K_def}
2 \alpha K_{ij} = D_i \beta_j + D_j \beta_i - {2\over 3} \gamma_{ij}
D_k \beta^k.
\end{equation}
Note that $\partial_t\gamma\neq 0$ unless $D_i\beta^i=0$.

To solve the Hamiltonian constraint equation, it is convenient to use
a conformal decomposition of the spatial metric.  To satisfy the
demand that the left-hand side of Eq.~(\ref{eqn:met_evol_tf}) be zero,
we choose the metric to be conformally flat\cite{Note2} so that
$\gamma^{-1/3}\gamma_{ij} = f_{ij}$, where $f_{ij}$ is the flat metric
in whatever coordinate system is used.  Therefore, we decompose the
spatial metric as
\begin{equation}
\gamma_{ij} = \Phi^4 f_{ij}.
\end{equation}
The conformal factor $\Phi$ is determined then by the
Hamiltonian constraint
\begin{equation}
\label{eqn:ham_const}
\nabla^2 \Phi = - {1 \over 8} \Phi^5 K^{ij} K_{ij} - 2 \pi \Phi^5 \rho,
\end{equation}
where the source term is
\begin{equation}
\label{eqn:rho_def}
\rho = n^a n^b T_{ab}.
\end{equation}
Here $n^a$ is the normal vector to a $t={}$constant slice, $T_{ab}$ is
the stress-energy tensor, and $\nabla^2$ is the flat-space
Laplacian. Note that although indices $i,j,\ldots$ range over
$1,\ldots,3$, indices $a,b,\ldots$ range over $0,\ldots,3$.

The shift vector is determined by substituting Eq.~(\ref{eqn:K_def})
into the momentum constraint
\begin{equation}
\label{eqn:mom_const}
D_j K^{ij} = 8 \pi S^i,
\end{equation}
where
\begin{equation}
\label{eqn:S_def}
S^a = - \gamma^a{}_b n_c T^{bc}.
\end{equation}
We use the results that for a conformally flat metric we may write
\begin{eqnarray}
D^j \beta^i + D^i \beta^j - {2 \over 3} \gamma^{ij} D_k \beta^k =
\hspace{1.25in} &
\nonumber \\
\Phi^{-4} \left[\nabla^j \beta^i + \nabla^i \beta^j -
{2 \over 3} f^{ij} \nabla_k \beta^k \right],
\end{eqnarray}
and for $K=0$,
\begin{equation}
D_j K^{ij} = \Phi^{-10} \nabla_j (\Phi^{10} K^{ij}),
\end{equation}
where $\nabla_j$ denotes the covariant derivative in flat space. Thus
Eq.~(\ref{eqn:mom_const}) becomes
\begin{eqnarray}
\nabla^2 \beta^i + {1 \over 3} \nabla^i (\nabla_j \beta^j) =
16 \pi \alpha \Phi^4 S^i \hspace{1.05in} & \\
\mbox{} + \left({1 \over \alpha} \nabla_j \alpha -
	{6 \over \Phi} \nabla_j \Phi \right) \!\!\!\left(
\nabla^j \beta^i + \nabla^i \beta^j - {2 \over 3}
f^{ij} \nabla_k \beta^k \right). \hspace{-0.25in}\nonumber
\end{eqnarray}
This equation can be simplified to two equations, one involving a
vector Laplacian and the other a scalar Laplacian, by setting
\begin{equation}
\label{eqn:shift_dec}
\beta^i = G^i - {1 \over 4} \nabla^i B.
\end{equation}
Then the two equations that must be solved are
\begin{eqnarray}
\label{eqn:G_eqn}
\nabla^2 G^i = 16 \pi \alpha \Phi^4 S^i \hspace{1.85in} & \\
\mbox{} + \left({1 \over \alpha} \nabla_j
\alpha - {6 \over \Phi} \nabla_j \Phi \right) \left(
\nabla^j \beta^i + \nabla^i \beta^j - {2 \over 3}
f^{ij} \nabla_k \beta^k \right) \hspace{-0.25in} \nonumber
\end{eqnarray}
and
\begin{equation}
\label{eqn:B_eqn}
\nabla^2 B = \nabla_i G^i.
\end{equation}

Though we are not imposing the full set of dynamical equations for the
evolution of $K_{ij}$, we do have the freedom to preserve the maximal
slicing condition (\ref{eqn:max_slice}) by requiring $\partial_t K=0$.
The resulting equation can also be written with a simple Laplacian by
using Eq.~(\ref{eqn:ham_const}).  The result is the lapse equation
\begin{equation}
\label{eqn:lapse_eqn}
\nabla^2 (\alpha \Phi) = (\alpha \Phi) \left[{7 \over 8}
\Phi^4 K_{ij} K^{ij} + 2 \pi \Phi^4 (\rho + 2S) \right],
\end{equation}
where
\begin{equation}
\label{eqn:Strace_def}
S=\gamma^{ij} T_{ij}.
\end{equation}

The above field equations, in combination with the matter equations to
be discussed below, form a coupled nonlinear set that must be solved
by iteration.  The boundary conditions for the field quantities follow
from asymptotic flatness; the specific form depends on the
application. We are especially interested in uniformly rotating
configurations such as binary neutron stars in synchronous orbit. For
such systems we work in a corotating coordinate system so that there
is no time variation of the fields (in the near equilibrium
approximation of the method).  Following Wilson\cite{Texas_90:Wilson},
we can implement this by replacing Eq.~(\ref{eqn:shift_dec}) with
\begin{equation}
\beta^i = G^i - {1 \over 4} \nabla^i B + \left({\bf \Omega} \times {\bf r}
 \right)^i,
\end{equation}
which leaves Eqs.~(\ref{eqn:B_eqn}) and (\ref{eqn:G_eqn}) unchanged.
Here $\bf \Omega$ is the constant angular velocity of the system.

For the matter, we will consider a perfect fluid for which
\begin{equation}
T_{ab} = \left(\rho_0 + \rho_i + P\right) U_a U_b + Pg_{ab}.
\end{equation}
Here $\rho_0$ is the rest-mass density, $\rho_i$ is the internal
energy density, $P$ is the pressure, and $U^a$ is the fluid
4-velocity.  For this source, the density $\rho$ in
Eq.~(\ref{eqn:rho_def}) is
\begin{equation}
\rho = \left(\rho_0 + \rho_i + P\right) \left(\alpha U^t\right)^2 - P,
\end{equation}
the momentum source $S^i$ in Eq.~(\ref{eqn:S_def}) is
\begin{equation}
S^i = \left(\rho_0 + \rho_i + P \right) \left(\alpha U^t\right) \gamma^{ij}
 U_j,
\end{equation}
and the source term $S$ in Eq.~(\ref{eqn:Strace_def}) is
\begin{equation}
S= \left(\rho_0 + \rho_i + P \right) \left[(\alpha U^t)^2 -1 \right] + 3P.
\end{equation}
We treat fluids that
are in uniform rotation, for which the 4-velocity $U^a$ is given by
\begin{equation}
\label{eqn:U_def}
 \vec{U} = U^t \left({\partial \over \partial t} + \Omega
{\partial \over \partial \phi} \right).
\end{equation}
The normalization condition $\vec U\cdot\vec U=-1$ gives
\begin{equation}
\alpha U^t = \left(1 + \Phi^{-4} f^{ij} U_i U_j \right)^{1/2}.
\end{equation}

Now consider the equations for the matter in the near equilibrium
approximation. The key approximation is that in the corotating frame
there is a Killing vector that is timelike everywhere. In the
nonrotating coordinates, this vector can be written as
\begin{equation}
\vec\xi = {\partial \over \partial t} + \Omega
{\partial \over \partial \phi}.
\end{equation}
Because the 4-velocity (\ref{eqn:U_def}) is proportional to a Killing
vector, the matter equations may be integrated to give the hydrostatic
equilibrium result\cite{problembook}
\begin{equation}
\label{eqn:enth_def}
{U^t \over h} = {\rm constant},
\end{equation}
where
\begin{equation}
\ln h \equiv \int {dP \over \rho_0 + \rho_i + P}.
\end{equation}
For a polytropic equation of state
\begin{equation}
P = K\rho^\Gamma_0,
\end{equation}
where $K$ and $\Gamma$ are constants,
we have
\begin{equation}
\rho_i = {P \over \Gamma - 1},\qquad
h = {\rho_0 + \rho_i + P \over \rho_0}.
\end{equation}
In this approximation, we have reduced all of the hydrodynamics to a
single algebraic equation, Eq.~(\ref{eqn:enth_def}).

\section{Axisymmetric Rotating Star: Equations}
\label{sec:axi_eqns}
To calibrate the method, we apply it to a true equilibrium system in
axisymmetry and compare with the complete numerical solution found
with no approximations. For this purpose, we use models of rotating
neutron stars supported by a polytropic equation of state. Fully
relativistic models have been constructed by several authors (see
Refs.~\cite{cook92,cook94a,cook94c} and references therein).  Solving
Einstein's equations for these stars is nontrivial numerically.  It is
only the recent availability of such solutions that makes this
calibration feasible.

In spherical polar coordinates and axisymmetry, we find that
Eqs.~(\ref{eqn:G_eqn}) and (\ref{eqn:B_eqn}) are satisfied by setting
the quantity $B$ of Eq.~(\ref{eqn:shift_dec}) to zero and with the
only nonzero component of the shift vector $\beta^\phi\equiv\beta$.
Note that this implies, not only that the left-hand side of
Eq.~(\ref{eqn:met_evol_tf}) is zero, but also that $\partial_t\gamma =
0$.  This means that we are finding a stationary solution of the
approximate equations.  Given this solution for the shift vector, the
term $K_{ij}K^{ij}$ appearing in Eqs.~(\ref{eqn:ham_const}) and
(\ref{eqn:lapse_eqn}) is given by
\begin{equation}
K_{ij} K^{ij} = {\sin^2 \theta \over 2 \alpha^2} \left(r^2 \beta^2_{,r} +
 \beta^2_{,\theta}\right),
\end{equation}
where commas denote partial derivatives.  Only the $\phi$-component of
the vector Eq.~(\ref{eqn:G_eqn}) is nontrivial, and becomes the scalar
equation
\begin{eqnarray}
\label{eqn:shift_eqn}
\left[\nabla^2 + {2 \over r} {\partial \over \partial r} +
{2 \cot \theta \over r^2} {\partial \over \partial \theta}
\right] \!\beta =
\left( {1 \over \alpha} {\partial \alpha \over \partial r} -
{6 \over \Phi}{\partial \Phi \over \partial r} \right)
\!{\partial \beta \over  \partial r} & \nonumber \\
\mbox{} + {1 \over r^2}
\left({1 \over \alpha} {\partial \alpha \over \partial \theta} -
{6 \over \Phi} {\partial \Phi \over \partial \theta} \right)
{\partial \beta \over \partial \theta} +
{16 \pi \alpha \over r^2 \sin^2 \theta} S_{\phi}.
\end{eqnarray}
The 4-velocity components appearing in the matter sources are given by
\begin{eqnarray}
U^t &=&  \left[\alpha^2 - \Phi^4 r^2 \sin^2 \theta
(\beta + \Omega)^2 \right]^{-1/2}, \\ \nonumber
U_\phi &=& \Phi^4 r^2 \sin^2 \theta\, U^t \left(\beta + \Omega \right).
\end{eqnarray}

The above equations turn out to be simplified versions of the exact
equations for stationary, axisymmetric configurations given in
Ref.~\cite{cook92} (henceforth CST\cite{Note3}).  The exact metric has
four nonzero metric coefficients, denoted by $\gamma$, $\rho$,
$\alpha$ and $\omega$ in CST, though the approximate metric here has
only three: $\alpha$, $\beta$, and $\Phi$. Thus even though there is
no dynamics in the field, and even though the equation of hydrostatic
equilibrium for the matter is rigorously obeyed, the Wilson scheme is
still an approximation for this problem. The correspondence between
the approximate and exact metric coefficients is given by
\begin{eqnarray}
\label{eqn:CST_coefs}
\alpha^2 &=& e^{\gamma + \rho}, \\ \nonumber
\Phi^4 &=& e^{\gamma - \rho} = e^{2 \alpha_{\rm CST}}, \\
\beta &=& - \omega. \nonumber
\end{eqnarray}
The fluid velocity $v$ in the ZAMO frame used in CST is given by
\begin{equation}
(\alpha U^t)^2 = {1 \over 1 - v^2}.
\end{equation}
In spherical symmetry, the approximate scheme reduces to the exact
scheme, with two nonzero metric coefficients.  We will now quantify
the degree of error in the nonspherical axisymmetric case.

We can take over the numerical scheme of CST to solve the approximate
equilibrium equations. In fact, the structure of the equations is very
close in that they involve the same differential operators on the
left-hand sides.  In particular, Eqs.~(\ref{eqn:ham_const}) and
(\ref{eqn:lapse_eqn}) involve $\nabla^2$, as in Eq.~(3) of CST, and
the operator in Eq.~(\ref{eqn:shift_eqn}) is the same as that in
Eq.~(5) of CST. Therefore the solution is computed as in Eqs.~(27) and
(29) of CST. The nondimensionalized source terms analogous to Eq.~(30)
of CST are
\begin{eqnarray}
\tilde{S}_\Phi(s,\mu) = - {1 \over 16}
{\Phi^7 \over \left(\alpha \Phi \right)^2} \left(1-\mu^2 \right)
\left({s \over 1-s}\right)^2 \hspace{0.55in} &\\
\mbox{} \times \Bigl\{\left[
\left(1-s\right) s \hat \omega_{,s} \right]^2
+ \left(1 - \mu^2 \right) \hat\omega^2_{,\mu}\Bigr\} \nonumber \\
\mbox{} - 2 \pi \Phi^5  \bar{r}_e^2 \left({s \over 1-s}\right)^2
\left[\left(\bar\rho_0 +\bar\rho_i + \bar{P} \right)
 {1 \over 1 -v^2} - \bar{P}\right], \hspace{-0.15in} \nonumber
\end{eqnarray}
\begin{eqnarray}
\tilde{S}_{\alpha\Phi}(s,\mu) = \alpha\Phi
 \Bigg[\!\!\Bigg[
{7 \over 16} {\Phi^6 \over (\alpha \Phi)^2} \left(1 - \mu^2\right)
\left({s \over 1-s}\right)^2 \hspace{0.26in} & \\
\mbox{}\times\Bigl\{[\left(1-s\right) s \hat\omega_{,s}]^2 +
\left(1-\mu^2\right)\hat\omega^2_{,\mu} \Bigr\}  \nonumber \\
\mbox{} + 2 \pi \Phi^4 \bar{r}^2_e \left({s\over 1-s}\right)^2
 \left[\left(\bar\rho_0 + \bar\rho_i +  \bar{P} \right)
 {1 \over 1-v^2} - \bar{P} \right] \hspace{-0.17in}\nonumber \\
\mbox{} + 4 \pi \Phi^4 \bar{r}^2_e \left({s \over 1-s}\right)^{\!2}
 \left[\left(\bar\rho_0 + \bar\rho_i +\bar{P}\right)
 {v^2 \over 1-v^2} + 3\bar{P}
\right]\!\!\Bigg]\!\!\Bigg], \hspace{-0.3in} \nonumber
\end{eqnarray}
and the source term analogous to Eq.~(32) of CST is
\begin{eqnarray}
\tilde{S}_{\hat\omega}(s,\mu) &=& s^2 (1-s)^2
\left[{1 \over \alpha\Phi} \left(\alpha \Phi\right)_{,s} - {7 \over \Phi}
\Phi_{,s} \right] \hat\omega_{,s} \nonumber \\
&&\mbox{} + (1 - \mu^2) \left[{1 \over \alpha \Phi}
\left(\alpha \Phi\right)_{,\mu} -
{7 \over \Phi} \Phi_{, \mu} \right] \hat\omega_{, \mu} \\ \nonumber
&&\mbox{} - 16 \pi \Phi^4 \bar{r}^2_e \left({s \over 1-s}\right)^2
 { \bar\rho_0 + \bar\rho_i+\bar{P}
\over 1-v^2} \left(\hat\Omega - \hat\omega \right),
\end{eqnarray}
where $s$ is an auxiliary radial coordinate defined in CST.  The
entire iterative scheme used to solve the approximate equations is
identical to the one in CST.

To calibrate the approximation, we first compute an exact sequence of
constant rest mass polytropes of increasing angular momentum. Each
member of the sequence is specified by two parameters: the ratio of
polar to equatorial radius, and the central rest-mass density. We next
compute the approximate sequence using the same values for these two
parameters for each model.  We then compare the metric coefficients of
corresponding models using the relationships in
(\ref{eqn:CST_coefs}). We also compare global quantities such as the
total mass and angular momentum. As a further diagnostic, we calculate
two relativistic virial quantities\cite{gourgoulhon94,bonazzola94}
whose values should be identically one for an exact equilibrium
solution.  In the notation of CST, these quantities are
\begin{eqnarray}
\lambda_{2d} = 32\pi\int\left[P + (\epsilon + P){v^2\over 1-v^2}\right]
e^{2\alpha} r dr d \theta \Bigg/ \hspace{0.25in} &\\
\int\Biggl\{\left({\partial\gamma\over\partial r} +
{\partial\rho\over\partial r}\right)^2 + {1 \over r^2}
\left({\partial\gamma\over\partial\theta} +
{\partial\rho\over\partial\theta}\right)^2 \hspace{0.7in}\nonumber \\
\mbox{} -  3e^{-2\rho}\sin^2\theta\left[
r^2 \left({\partial\omega\over\partial r}\right)^2 +
\left({\partial\omega\over\partial theta}\right)^2 \right]\Biggr\}
r dr d\theta, \hspace{-0.3in}\nonumber
\end{eqnarray}
\begin{eqnarray}
\lambda_{3d} = 16 \pi \int \left[3P + (\epsilon + P )
{v^2 \over 1-v^2} \right] \hspace{0.85in} \\
\mbox{}\times
e^{2\alpha+(\gamma-\rho)/2} r^2 \sin\theta dr d\theta\Bigg/ &\nonumber \\
\int\bigg[\partial(\gamma + \rho)\partial(\gamma + \rho) -
\partial\alpha\partial\gamma + \partial\alpha\partial\rho
\hspace{0.75in} \nonumber \\
\mbox{} - {1\over 2r}(1-e^{2\alpha-\gamma+\rho})
\Bigl(4{\partial\alpha\over\partial r} +
{4 \over r \tan\theta}{\partial\alpha\over\partial\theta}
- {\partial\gamma\over partial r} \nonumber \nonumber \\
\mbox{} - {1 \over r \tan\theta}{\partial\gamma\over\partial\theta} +
{\partial\rho\over\partial r} + {1 \over r \tan\theta}
{\partial\rho\over\partial\theta}\Bigr) \hspace{-0.25in} \nonumber \\
\mbox{} -{3\over2}e^{-2\rho}r^2\sin^2\theta
\partial\omega\partial\omega\bigg]
e^{(\gamma-\rho)/2} r^2 \sin \theta dr d\theta,\hspace{-0.35in}\nonumber
\end{eqnarray}
where
\begin{equation}
\partial \alpha \partial \rho \equiv {\partial \alpha \over \partial r}
{\partial \rho \over \partial r} + {1 \over r^2} {\partial \alpha \over
\partial r} {\partial \rho \over \partial \theta}
\end{equation}
and $\epsilon = \rho_0 + \rho_i$ is the total mass-energy density.
Here $\lambda_{3d}$ involves an integration with a 3-dimensional
volume element $r^2\sin\theta\,dr\,d\theta$ and is the relativistic
generalization of the classical virial theorem
\begin{equation}
2 E_{\rm kin} + 3(\Gamma-1) E_{\rm int} + U_{\rm grav} =0.
\end{equation}
The quantity $\lambda_{2d}$ involves an integration with a
2-dimensional volume element $r\,dr\,d\theta$. The discrepancy from
unity is a measure of {\it numerical} error for our solutions of the
exact equations. It is a measure of the larger {\it inherent} error
for our solutions of the approximate equations.

\section{Axisymmetric Rotating Star: Numerical Results}
\label{sec:axi_results}
To calibrate the approximate scheme against the exact solution, we
choose the most stringent case, in which the configuration is very
relativistic and rapidly rotating. When it is rotating rapidly, there
are large deviations from spherical symmetry, so that the
approximation is no longer exact. For polytropes, the largest rotation
is attained for nearly incompressible matter, i.e. for large
$\Gamma=1+1/n$ or small polytropic index $n$. We choose $n=0.5$.

In constructing an exact sequence of rotating equilibria as a
benchmark, we start with a nonrotating star having a central value of
energy density $\bar\epsilon = 1$ (note that all ``barred'' quantities
are nondimensional as defined in CST).  This configuration is
relativistic, with $M/R=0.298$ and rest mass $\bar{M}_0 = 0.148$, just
below the maximum rest mass of a nonrotating star for this equation of
state ($\bar{M}_0 = 0.151$).  Holding the rest mass constant, we
construct a sequence of increasing uniform rotation, up to the point
of mass shedding.  As described above, we then construct the
corresponding models with the same central value of $\bar\epsilon$ and
ratio of polar to equatorial radius using the approximate scheme. A
comparison of some of the global quantities for the sequence is given
in Table~\ref{tab:Normal_table}. The high values of polar redshift
$Z_p$ and $T/W$ confirm that the sequence is both highly relativistic
and rapidly rotating. As expected, the deviations are largest near the
mass shed limit, but even there they are never worse than about 1\%.

\begin{figure}
\epsfxsize=3.4in\epsffile{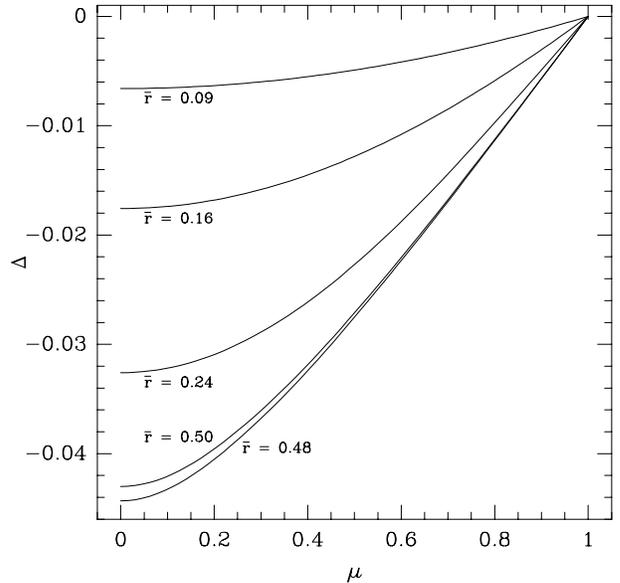}
\caption{Angular profile of the deviation of the exact solution from
conformal flatness at selected radii.  The deviation $\Delta$ is
defined in Eq.~(\protect\ref{eqn:Delta_def}).  The star is a rapidly
rotating, highly relativistic polytrope with $n=0.5$ and rest mass
just below the maximum rest mass of a nonrotating star for this
equation of state.  The radii $\bar{r}$ are in the nondimensional
units of CST, and $\mu=\cos\theta$.}
\label{fig:conf_flatness}
\end{figure}

We can understand why the overall discrepancy is small by looking at
Fig.~\ref{fig:conf_flatness}.  Here we plot a measure of the deviation
in the exact solution from conformal flatness, which is assumed in the
approximate method.  In the figure we plot the angular profile at
selected radii of the quantity
\begin{equation}
\label{eqn:Delta_def}
\Delta\equiv{\alpha_{\rm CST} - (\gamma - \rho)/2 \over \alpha_{\rm CST}}
\end{equation}
computed for the exact rotating model with $T/W = 0.159$. Note that
this quantity is identically zero on the axis because of local
flatness there.  The maximum deviation occurs on the equator
($\bar{r}=0.48$), but is only about 5\%.  Outside the star, $\Delta\to
0$ as $r\to\infty$.

\begin{figure}
\epsfxsize=3.4in\epsffile{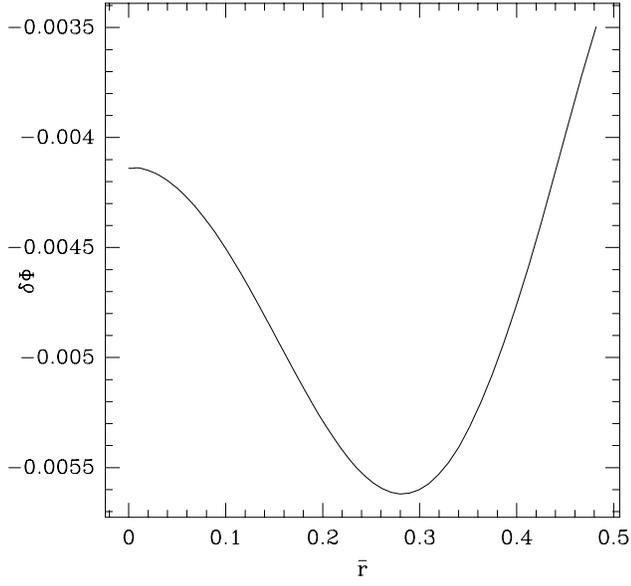}
\caption{Fractional error in the conformal factor $\Phi$ along an
equatorial radius for the star in
Fig.~\protect\ref{fig:conf_flatness}.}
\label{fig:phi_error}
\end{figure}

\begin{figure}
\epsfxsize=3.4in\epsffile{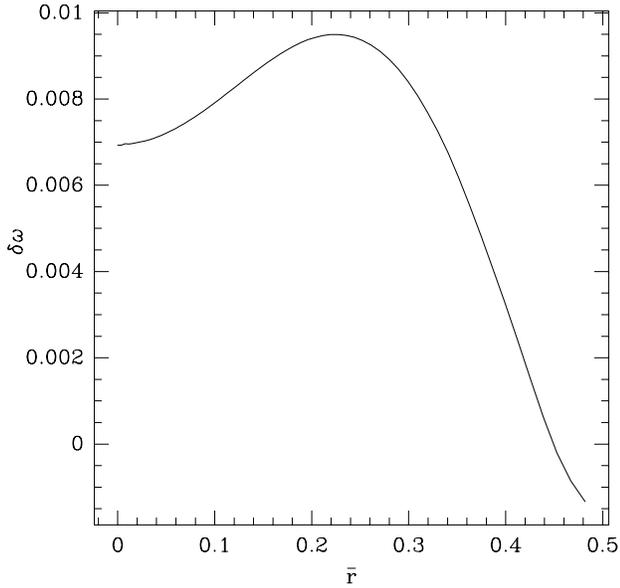}
\caption{Fractional error in the metric coefficient $\omega$ along an
equatorial radius for the star in
Fig.~\protect\ref{fig:conf_flatness}.}
\label{fig:omega_error}
\end{figure}

\begin{figure}
\epsfxsize=3.4in\epsffile{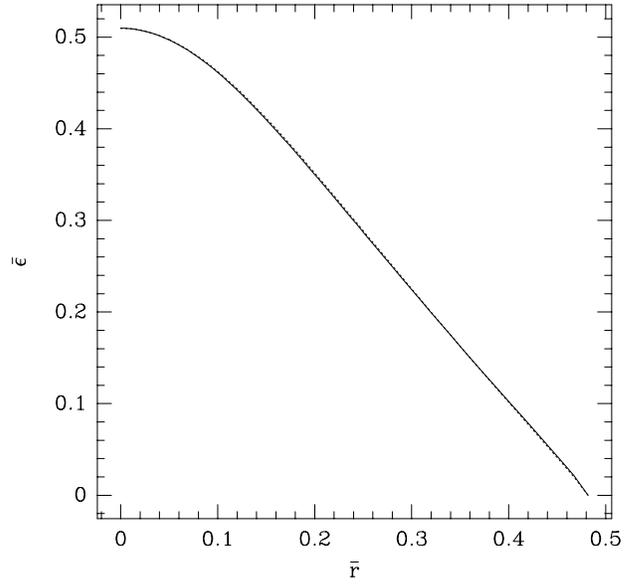}
\caption{Total mass-energy density $\bar\epsilon$ along an equatorial
radius for the star in Fig.~\protect\ref{fig:conf_flatness}. The solid
line shows the exact solution, the dotted line, the approximate
solution.}
\label{fig:density}
\end{figure}

In Fig.~\ref{fig:phi_error} we plot along an equatorial radius the
fractional error in the conformal factor,
\begin{equation}
\delta\Phi = {\Phi - \Phi_{\rm exact} \over \Phi_{\rm exact}},
\end{equation}
where $\Phi_{\rm exact}\equiv\exp[(\gamma-\rho)/4]$.  Similarly, in
Fig.~\ref{fig:omega_error} we plot the fractional error
$\delta\omega$.  Figure~\ref{fig:density} shows the mass-energy
$\bar\epsilon$ along an equatorial radius for the two schemes.  The
two profiles are almost coincident.

\begin{figure}
\epsfxsize=3.4in\epsffile{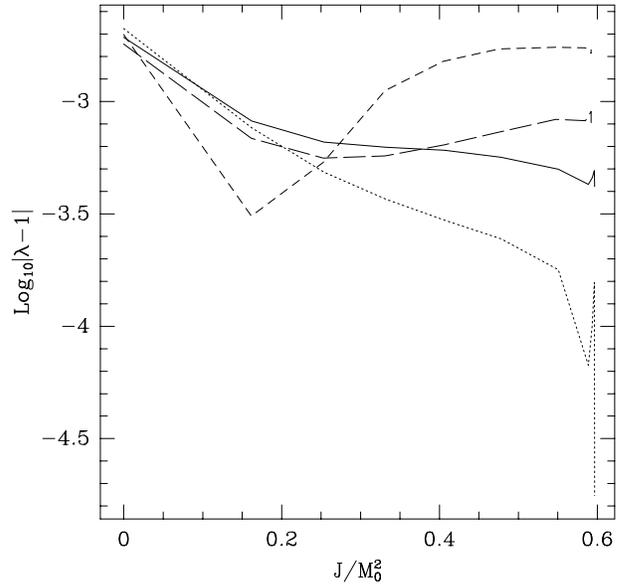}
\caption{Virial quantities along the sequence in
Table~\protect\ref{tab:Normal_table}. Results for the exact equations
are shown by the solid line for $\lambda_{2d}$ and the dotted line for
$\lambda_{3d}$.  Results for the approximation are shown by the short
dash line for $\lambda_{2d}$ and the long dash line for
$\lambda_{3d}$.}
\label{fig:virial_normal}
\end{figure}

A further comparison is provided by Fig.~\ref{fig:virial_normal},
which shows the virial quantities $\lambda_{2d}$ and $\lambda_{3d}$
along the sequence, computed for each of the two schemes. In the case
of the exact method, the deviation from unity is a measure of
numerical error, which is less than 0.1\%.  The deviation for the
approximate method measures the inherent error, which is about a
factor of 10 bigger.

\begin{figure}
\epsfxsize=3.4in\epsffile{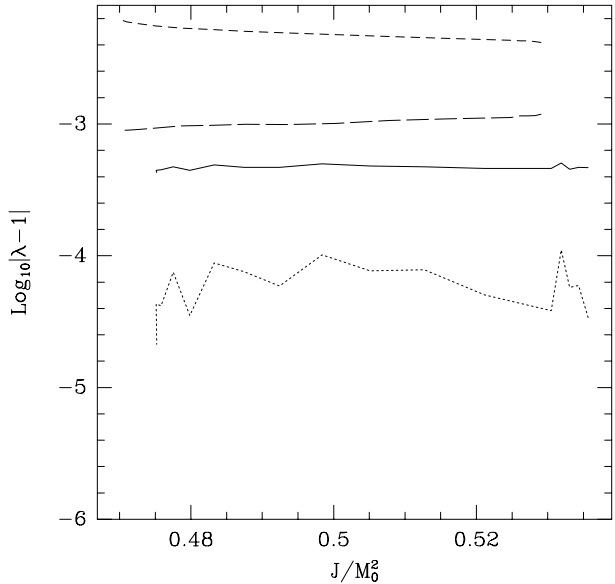}
\caption{Virial quantities along the supramassive sequence in
Table~\protect\ref{tab:Supra_table}.  Results for the exact equations
are shown by the solid line for $\lambda_{2d}$ and the dotted line for
$\lambda_{3d}$.  Results for the approximation are shown by the short
dash line for $\lambda_{2d}$ and the long dash line for
$\lambda_{3d}$.}
\label{fig:virial_supra}
\end{figure}

To push the approximate scheme to the limit, we now consider a second
equilibrium sequence, a ``supramassive'' sequence.  This sequence has
no nonrotating member, since its rest mass exceeds the maximum rest
mass of a nonrotating star for this equation of state ($\bar{M}_0 =
0.151$).  Thus the sequence exists only by virtue of rotation. We
construct the supramassive sequence with $\bar{M}_0 = 0.176$.  We
expect the discrepancy between the approximate and exact methods to be
somewhat larger for this sequence since it is everywhere far from
spherical symmetry.  This expectation is borne out in
Table~\ref{tab:Supra_table} and Fig.~\ref{fig:virial_supra}.
Nevertheless, the discrepancy is not very large.

\section{Conclusion}
\label{sec:conclusions}
We have tested Wilson's approximation scheme on rapidly rotating
relativistic stars. Since these are equilibrium objects, it is
necessary that the scheme give reasonably accurate results if we are
to believe its predictions for more complicated systems such as
binaries. In fact, we have found that the method works remarkably
well, even for highly relativistic objects far from spherical
symmetry. The largest errors in any quantities we examined were around
5\%, and in general the errors were much smaller. Global measures such
as virial quantities were in error by far less than 1\%. This
agreement is very encouraging.

\acknowledgments
We thank E. Gourgoulhon for useful discussions about the work on
relativistic virial quantities in
Refs.~\cite{gourgoulhon94,bonazzola94}.  This work was supported in
part by NSF Grants Nos.~AST 91-19475 and PHY~94-08378, and by NASA
Grant NAGW-2364 to Cornell University.  We also acknowledge support
from the Grand Challenge Grant NSF PHY 93-18152 / ASC 93-18152.


\begin{references}

\bibitem[*]{}
Departments of Astronomy and Physics, Cornell University.

\bibitem{Frontiers:Wilson}
J.~R.\ Wilson and G.~J.\ Mathews, in {\em Frontiers in Numerical
  Relativity}, edited by C.~R.\ Evans, L.~S.\ Finn, and D.~W.\ Hobill
  (Cambridge University Press, Cambridge, England, 1989), pp.\
  306--314.

\bibitem{Texas_90:Wilson}
J.~R.\ Wilson, in {\em Texas Symposium on 3-Dimensional Numerical
  Relativity}, edited by R.~A.\ Matzner (University of Texas, Austin,
  Texas, 1990).

\bibitem{Note1}
Other approximations for binary black holes have been presented by
Detweiler\cite{detweiler92} and Cook\cite{cook94e}.

\bibitem{Note2}
Note that we are free to choose $\gamma^{-1/3}\gamma_{ij}$ to be any
time-independent metric.  The choice of a flat metric simplifies the
derivations slightly and results in familiar flat-space differential
operators.

\bibitem{problembook}
A.~P.\ Lightman, W.~H. Press, R.~H.\ Price, and S.~A.\ Teukolsky, {\em
  Problem Book in Relativity and Gravitation} (Princeton University
  Press, Princeton, New Jersey, 1975).

\bibitem{cook92}
G.~B.\ Cook, S.~L.\ Shapiro, and S.~A.\ Teukolsky, Astrophys.\ J. {\bf
  398}, 203 (1992).

\bibitem{cook94a}
G.~B.\ Cook, S.~L.\ Shapiro, and S.~A.\ Teukolsky, Astrophys.\ J. {\bf
  422}, 227 (1994).

\bibitem{cook94c}
G.~B.\ Cook, S.~L.\ Shapiro, and S.~A.\ Teukolsky, Astrophys.\ J. {\bf
  424}, 823 (1994).

\bibitem{Note3}
Note that some minor errors in Ref.~\cite{cook92} were corrected in
Ref.~\cite{cook94a}.

\bibitem{gourgoulhon94}
E.\ Gourgoulhon and S.\ Bonazzola, Class.\ Quantum\ Gravit. {\bf 11},  443
  (1994).

\bibitem{bonazzola94}
S.\ Bonazzola and E.\ Gourgoulhon, Class.\ Quantum\ Gravit. {\bf 11},
1775 (1994).

\bibitem{detweiler92}
J.~K.\ Blackburn and S. Detweiler, Phys.\ Rev.\ D {\bf 46}, 2318
(1992).

\bibitem{cook94e}
G.~B.\ Cook, Phys.\ Rev.\ D {\bf 50}, 5025 (1994).

\end{references}

\onecolumn
%
%
\mediumtext
\begin{table}
\caption{Quantities characterizing a ``normal'' evolutionary sequence
of n=0.5 polytropic neutron star models.  For each value of the
central energy density $\bar\epsilon$ we first display the results
from solving the exact equations.  Below this, indicated with a dash
in the energy density column, are the results obtained by solving the
approximate equations.}
\label{tab:Normal_table}
\begin{tabular}{ccccccccccc}
$\bar\epsilon_c$\tablenote{Central energy density.} &
$\bar\Omega$\tablenote{Angular velocity measured at infinity.} &
$\bar{I}$\tablenote{Moment of inertia.} &
$\bar{M}$\tablenote{Total mass-energy.} &
$\bar{M}_0$\tablenote{Rest mass.} &
$T/W$\tablenote{Rotational kinetic energy over gravitational binding
	energy.} &
$\bar{R}_e$\tablenote{Circumferential radius.} &
$e$\tablenote{Eccentricity.} &
$\omega_c/\Omega_c$\tablenote{Measure of frame dragging.} &
$v/c$\tablenote{Matter velocity at equator.} &
$Z_p$\tablenote{Polar redshift.} \\
\tableline
1.0000 & 0.0000 & 0.01014 & 0.1232 & 0.1484 & 0.0000 & 0.4137 &
	0.000 & 0.7482 & 0.0000 & 0.5726 \\
 --- & 0.0000 & 0.01014 & 0.1232 & 0.1484 & 0.0000 & 0.4137 &
	0.000 & 0.7480 & 0.0000 & 0.5725 \\
0.9029 & 0.3339 & 0.01069 & 0.1237 & 0.1484 & 0.0150 & 0.4274 &
	0.271 & 0.7291 & 0.1594 & 0.5660 \\
 --- & 0.3332 & 0.01068 & 0.1237 & 0.1484 & 0.0149 & 0.4270 &
	0.275 & 0.7289 & 0.1590 & 0.5658 \\
0.8152 & 0.4886 & 0.01144 & 0.1245 & 0.1484 & 0.0360 & 0.4454 &
	0.418 & 0.7115 & 0.2420 & 0.5641 \\
 --- & 0.4881 & 0.01141 & 0.1245 & 0.1484 & 0.0358 & 0.4445 &
	0.423 & 0.7115 & 0.2413 & 0.5638 \\
0.7360 & 0.5923 & 0.01237 & 0.1255 & 0.1484 & 0.0600 & 0.4674 &
	0.532 & 0.6946 & 0.3064 & 0.5623 \\
 --- & 0.5927 & 0.01231 & 0.1254 & 0.1484 & 0.0597 & 0.4658 &
	0.537 & 0.6947 & 0.3057 & 0.5620 \\
0.6645 & 0.6618 & 0.01352 & 0.1265 & 0.1484 & 0.0860 & 0.4942 &
	0.627 & 0.6780 & 0.3604 & 0.5589 \\
 --- & 0.6632 & 0.01344 & 0.1265 & 0.1484 & 0.0858 & 0.4920 &
	0.632 & 0.6782 & 0.3598 & 0.5587 \\
0.6000 & 0.7040 & 0.01496 & 0.1276 & 0.1484 & 0.1134 & 0.5279 &
	0.707 & 0.6612 & 0.4081 & 0.5526 \\
 --- & 0.7064 & 0.01485 & 0.1276 & 0.1485 & 0.1132 & 0.5252 &
	0.712 & 0.6616 & 0.4076 & 0.5527 \\
0.5417 & 0.7224 & 0.01680 & 0.1287 & 0.1484 & 0.1416 & 0.5743 &
	0.778 & 0.6440 & 0.4552 & 0.5422 \\
 --- & 0.7256 & 0.01665 & 0.1288 & 0.1485 & 0.1413 & 0.5710 &
	0.782 & 0.6445 & 0.4550 & 0.5426 \\
0.5148 & 0.7226 & 0.01795 & 0.1293 & 0.1484 & 0.1559 & 0.6124 &
	0.817 & 0.6349 & 0.4875 & 0.5348 \\
 --- & 0.7260 & 0.01777 & 0.1293 & 0.1485 & 0.1555 & 0.6091 &
	0.820 & 0.6354 & 0.4876 & 0.5351 \\
0.5115 & 0.7221 & 0.01812 & 0.1294 & 0.1484 & 0.1577 & 0.6217 &
	0.824 & 0.6337 & 0.4956 & 0.5337 \\
 --- & 0.7254 & 0.01792 & 0.1294 & 0.1485 & 0.1572 & 0.6184 &
	0.827 & 0.6341 & 0.4958 & 0.5339 \\
0.5098 & 0.7218 & 0.01821 & 0.1294 & 0.1484 & 0.1587 & 0.6303 &
	0.830 & 0.6331 & 0.5034 & 0.5332 \\
 --- & 0.7249 & 0.01799 & 0.1294 & 0.1484 & 0.1581 & 0.6272 &
	0.833 & 0.6334 & 0.5037 & 0.5331 \\
0.5094 & 0.7216 & 0.01822 & 0.1294 & 0.1484 & 0.1588 & 0.6331 &
	0.832 & 0.6329 & 0.5060 & 0.5330 \\
 --- & 0.7248 & 0.01800 & 0.1294 & 0.1484 & 0.1581 & 0.6300 &
	0.835 & 0.6332 & 0.5063 & 0.5328 \\
0.5094 & 0.7216 & 0.01822 & 0.1294 & 0.1484 & 0.1589 & 0.6332 &
	0.832 & 0.6329 & 0.5061 & 0.5330 \\
 --- & 0.7248 & 0.01800 & 0.1294 & 0.1484 & 0.1581 & 0.6302 &
	0.835 & 0.6332 & 0.5065 & 0.5328 \\
0.5094 & 0.7216 & 0.01822 & 0.1294 & 0.1484 & 0.1589 & 0.6333 &
	0.832 & 0.6329 & 0.5062 & 0.5330 \\
 --- & 0.7248 & 0.01800 & 0.1294 & 0.1484 & 0.1581 & 0.6302 &
	0.835 & 0.6332 & 0.5066 & 0.5328 \\
\end{tabular}
\end{table}

\begin{table}
\caption{Quantities characterizing a ``supramassive'' evolutionary
sequence of n=0.5 polytropic neutron star models.  Entries are as
described for Table~\protect\ref{tab:Normal_table}.}
\label{tab:Supra_table}
\begin{tabular}{ccccccccccc}
$\bar\epsilon_c$\tablenote{Central energy density.} &
$\bar\Omega$\tablenote{Angular velocity measured at infinity.} &
$\bar{I}$\tablenote{Moment of inertia.} &
$\bar{M}$\tablenote{Total mass-energy.} &
$\bar{M}_0$\tablenote{Rest mass.} &
$T/W$\tablenote{Rotational kinetic energy over gravitational binding
	energy.} &
$\bar{R}_e$\tablenote{Circumferential radius.} &
$e$\tablenote{Eccentricity.} &
$\omega_c/\Omega_c$\tablenote{Measure of frame dragging.} &
$v/c$\tablenote{Matter velocity at equator.} &
$Z_p$\tablenote{Polar redshift.} \\
\tableline
1.0957 & 0.9464 & 0.01552 & 0.1471 & 0.1758 & 0.1248 & 0.4819 &
	0.703 & 0.8344 & 0.5196 & 0.9282 \\
 --- & 0.9549 & 0.01525 & 0.1472 & 0.1760 & 0.1246 & 0.4760 &
	0.714 & 0.8355 & 0.5192 & 0.9294 \\
1.0602 & 0.9362 & 0.01571 & 0.1472 & 0.1758 & 0.1254 & 0.4857 &
	0.705 & 0.8284 & 0.5174 & 0.9132 \\
 --- & 0.9444 & 0.01544 & 0.1472 & 0.1760 & 0.1251 & 0.4799 &
	0.717 & 0.8295 & 0.5170 & 0.9144 \\
1.0258 & 0.9269 & 0.01592 & 0.1472 & 0.1758 & 0.1264 & 0.4901 &
	0.709 & 0.8223 & 0.5163 & 0.8989 \\
 --- & 0.9349 & 0.01566 & 0.1473 & 0.1760 & 0.1261 & 0.4843 &
	0.720 & 0.8234 & 0.5158 & 0.9001 \\
0.9925 & 0.9182 & 0.01615 & 0.1473 & 0.1758 & 0.1277 & 0.4948 &
	0.714 & 0.8162 & 0.5158 & 0.8851 \\
 --- & 0.9260 & 0.01588 & 0.1473 & 0.1760 & 0.1274 & 0.4892 &
	0.724 & 0.8172 & 0.5153 & 0.8861 \\
0.9603 & 0.9105 & 0.01640 & 0.1474 & 0.1758 & 0.1294 & 0.5003 &
	0.720 & 0.8100 & 0.5166 & 0.8719 \\
 --- & 0.9182 & 0.01614 & 0.1474 & 0.1760 & 0.1291 & 0.4946 &
	0.730 & 0.8110 & 0.5162 & 0.8730 \\
0.9292 & 0.9031 & 0.01668 & 0.1475 & 0.1758 & 0.1314 & 0.5062 &
	0.726 & 0.8037 & 0.5179 & 0.8591 \\
 --- & 0.9107 & 0.01641 & 0.1476 & 0.1760 & 0.1311 & 0.5005 &
	0.736 & 0.8048 & 0.5176 & 0.8602 \\
0.8991 & 0.8964 & 0.01698 & 0.1476 & 0.1758 & 0.1337 & 0.5127 &
	0.734 & 0.7974 & 0.5203 & 0.8467 \\
 --- & 0.9038 & 0.01671 & 0.1477 & 0.1760 & 0.1334 & 0.5071 &
	0.743 & 0.7985 & 0.5200 & 0.8478 \\
0.8699 & 0.8901 & 0.01731 & 0.1478 & 0.1758 & 0.1365 & 0.5202 &
	0.742 & 0.7911 & 0.5239 & 0.8348 \\
 --- & 0.8975 & 0.01703 & 0.1479 & 0.1760 & 0.1362 & 0.5145 &
	0.751 & 0.7922 & 0.5237 & 0.8361 \\
0.8417 & 0.8839 & 0.01766 & 0.1480 & 0.1758 & 0.1395 & 0.5284 &
	0.752 & 0.7847 & 0.5283 & 0.8231 \\
 --- & 0.8912 & 0.01738 & 0.1481 & 0.1760 & 0.1392 & 0.5228 &
	0.760 & 0.7858 & 0.5282 & 0.8244 \\
0.8144 & 0.8779 & 0.01805 & 0.1482 & 0.1758 & 0.1429 & 0.5380 &
	0.763 & 0.7783 & 0.5343 & 0.8117 \\
 --- & 0.8852 & 0.01776 & 0.1483 & 0.1760 & 0.1426 & 0.5324 &
	0.771 & 0.7794 & 0.5343 & 0.8129 \\
0.7880 & 0.8719 & 0.01847 & 0.1484 & 0.1758 & 0.1466 & 0.5493 &
	0.775 & 0.7719 & 0.5423 & 0.8004 \\
 --- & 0.8790 & 0.01817 & 0.1485 & 0.1759 & 0.1462 & 0.5438 &
	0.782 & 0.7729 & 0.5424 & 0.8015 \\
0.7625 & 0.8654 & 0.01894 & 0.1487 & 0.1758 & 0.1506 & 0.5645 &
	0.790 & 0.7654 & 0.5546 & 0.7891 \\
 --- & 0.8723 & 0.01862 & 0.1487 & 0.1759 & 0.1500 & 0.5591 &
	0.797 & 0.7663 & 0.5549 & 0.7898 \\
0.7593 & 0.8648 & 0.01901 & 0.1487 & 0.1758 & 0.1512 & 0.5677 &
	0.793 & 0.7646 & 0.5579 & 0.7878 \\
 --- & 0.8716 & 0.01868 & 0.1487 & 0.1759 & 0.1506 & 0.5623 &
	0.800 & 0.7655 & 0.5583 & 0.7884 \\
0.7562 & 0.8637 & 0.01907 & 0.1487 & 0.1758 & 0.1517 & 0.5701 &
	0.795 & 0.7637 & 0.5598 & 0.7863 \\
 --- & 0.8705 & 0.01873 & 0.1487 & 0.1759 & 0.1510 & 0.5647 &
	0.802 & 0.7646 & 0.5602 & 0.7867 \\
0.7531 & 0.8629 & 0.01914 & 0.1488 & 0.1758 & 0.1522 & 0.5740 &
	0.799 & 0.7629 & 0.5639 & 0.7849 \\
 --- & 0.8695 & 0.01878 & 0.1487 & 0.1758 & 0.1514 & 0.5687 &
	0.805 & 0.7637 & 0.5644 & 0.7851 \\
0.7500 & 0.8620 & 0.01921 & 0.1488 & 0.1758 & 0.1528 & 0.5818 &
	0.806 & 0.7621 & 0.5729 & 0.7835 \\
 --- & 0.8682 & 0.01881 & 0.1487 & 0.1757 & 0.1517 & 0.5767 &
	0.812 & 0.7627 & 0.5736 & 0.7828 \\
\end{tabular}
\end{table}

\end{document}